\documentclass[final,5p,times,twocolumn]{elsarticle}
\usepackage{lineno}

\usepackage{graphicx}

\usepackage{amssymb}

\usepackage{amsmath}
\usepackage[hyphens]{url}

\usepackage{todonotes}

\usepackage{siunitx}

\usepackage{caption}
\biboptions{sort&compress}

\journal{Nuclear Instruments and Methods in Physics Research A}

\begin{document}

\begin{frontmatter}



%
%

\title{In-beam internal conversion electron spectroscopy with the SPICE detector}


\author[triumf]{M.~Moukaddam\fnref{fn1}}
\author[triumf]{J.~Smallcombe\corref{cor1}}
\author[triumf,surrey]{L.J.~Evitts}
\author[triumf]{A.B.~Garnsworthy}
\author[sfu]{C.~Andreoiu}
\author[triumf]{G.C.~Ball}
\author[triumf,ubc]{J.~Berean-Dutcher}
\author[triumf]{D.~Bishop}
\author[triumf,surrey]{C.~Bolton}
\author[triumf]{R.~Caballero-Folch}
\author[triumf]{M.~Constable}
\author[sfu]{D.S.~Cross}
\author[toronto]{T.E.~Drake}
\author[uoguelph]{R.~Dunlop}
\author[uoguelph]{P.E.~Garrett}
\author[triumf]{S.~Georges}
\author[triumf]{G.~Hackman}
\author[triumf,surrey]{S.~Hallam}
\author[triumf]{J.~Henderson\fnref{fn2}}
\author[triumf]{R.~Henderson}
\author[triumf,ubc]{R.~Kr\"{u}cken}
\author[triumf]{L.~Kurchaninov}
\author[triumf]{A.~Kurkjian}
\author[uoguelph]{B.~Olaizola\fnref{fn3}} 
\author[triumf]{E.~O'Sullivan}
\author[triumf]{P.~Lu}
\author[triumf,ubc]{J.~Park\fnref{fn4}}
\author[kentucky]{E.E.~Peters}
\author[sfu]{J.L.~Pore\fnref{fn5}} 
\author[uoguelph]{E.T.~Rand}
\author[triumf]{P.~Ruotsalainen\fnref{fn6}} 
\author[triumf]{J.K.~Smith\fnref{fn7}}
\author[triumf]{D.~Southall\fnref{fn8}} 
\author[triumf,surrey]{M.~Spencer}
\author[uoguelph]{C.E.~Svensson}
\author[triumf]{M.~Wiens}
\author[triumf,york]{M.~Williams}
\author[kentucky]{S.W.~Yates}
\author[uoguelph]{T.~Zidar}

\address[triumf]{TRIUMF, 4004 Wesbrook Mall, Vancouver, British Columbia, V6T 2A3 Canada}
\address[surrey]{Department of Physics, University of Surrey, Guildford, Surrey, GU2 7XH United Kingdom}
\address[sfu]{Department of Chemistry, Simon Fraser University, Burnaby, British Columbia, V5A 1S6 Canada}
\address[ubc]{Department of Physics and Astronomy, University of British Columbia, Vancouver, B.C., V6T 1Z1 Canada}
\address[toronto]{Department of Physics, University of Toronto, Toronto, Ontario, M5S 1A7 Canada}
\address[uoguelph]{Department of Physics, University of Guelph, Guelph, Ontario, N1G 2W1 Canada}
\address[york]{Department of Physics, University of York, Heslington, York, YO10 5DD United Kingdom}
\address[kentucky]{Departments of Chemistry and Physics \& Astronomy, University of Kentucky, Lexington, Kentucky 40506-0055 USA}

\cortext[cor1]{Corresponding Author: jsmallcombe@triumf.ca}
\fntext[fn1]{Present Address: Department of Physics, University of Surrey, Guildford, Surrey, GU2 7XH, UK}
\fntext[fn2]{Present Address: Lawrence Livermore National Laboratory, Livermore, CA 94550, USA}
\fntext[fn3]{Present Address: TRIUMF, 4004 Wesbrook Mall, Vancouver, British Columbia, V6T 2A3 Canada}
\fntext[fn4]{Present Address: Department of Physics, Lund University, 22100 Lund, Sweden}
\fntext[fn5]{Present Address: Lawrence Berkeley National Laboratory, Berkeley, CA 94720, USA}
\fntext[fn6]{Present Address: University of Jyv\"{a}skyl\"{a}, Department of Physics, P.O. Box 35, FI-40014 Jyv\"{a}skyl\"{a}, Finland}
\fntext[fn7]{Present Address: Department of Physics, Reed College, Portland, OR 97202, USA}
\fntext[fn8]{Present Address: Department of Physics, University of Chicago, Chicago, Illinois 60637, USA}

\begin{abstract}
The SPectrometer for Internal Conversion Electrons (SPICE) has been commissioned for use in conjunction with the TIGRESS $\gamma$-ray spectrometer at TRIUMF's ISAC-II facility. SPICE features a permanent rare-earth magnetic lens to collect and direct internal conversion electrons emitted from nuclear reactions to a thick, highly segmented, lithium-drifted silicon detector. This arrangement, combined with TIGRESS, enables in-beam $\gamma$-ray and internal conversion electron spectroscopy to be performed with stable and radioactive ion beams. Technical aspects of the device, capabilities, and initial performance are presented.
\end{abstract}

\begin{keyword}
Internal conversion electron\sep in-beam electron spectroscopy\sep shape coexistence

\end{keyword}

\end{frontmatter}


\section{Introduction}
\label{sec:introduction}
Radioactive ion-beam (RIB) facilities have permitted the study of nuclear systems far from stability, where subtle features of the nuclear force may become more prominent. The study of nuclei at the extremes of the nuclear landscape offers decisive tests of our understanding of the atomic nucleus.
Typically, however, the intensities of RIBs are below $1\times10^{6}$\,pps, significantly lower than those of stable beams. This necessitates the development of advanced detection systems with high efficiency and resolving power.

The $\gamma$-ray spectrometer TIGRESS~\cite{svensson_jpg31, hackman_hfi214} (TRIUMF-ISAC Gamma-Ray Escape Suppressed Spectrometer)  is one such device, adapted to the study of low-intensity RIBs.
Utilising segmented, high-efficiency, high-purity germanium (HPGe) clover detectors, TIGRESS is used to perform in-beam $\gamma$-ray spectroscopy to reveal nuclear structure (energy levels, electromagnetic transition strengths and multipolarities) in rare nuclei. Ancillary detectors (\textit{e.g.}~recoil analysers~\cite{davids_nim205}, charged-particle detectors~\cite{diget_jinst211} and recoil-distance devices~\cite{voss_nim214,Chester2018}) are crucial to enhance reaction channel selectivity and measurement precision or to access complementary experimental data.

\begin{figure*}[tb]
 \includegraphics[width=1.0\linewidth]{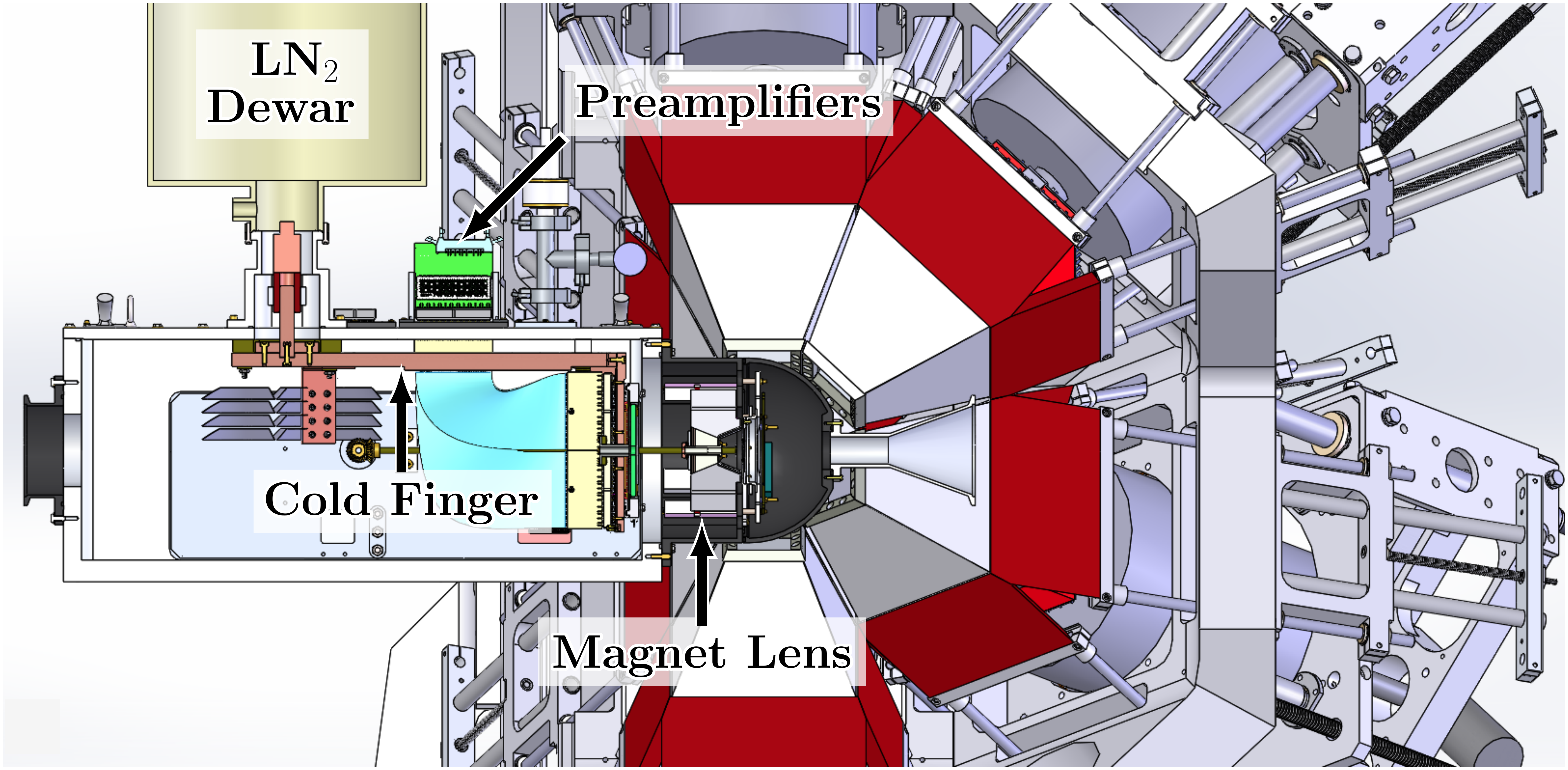}
\caption{Solidworks cross-section rendering of SPICE coupled with TIGRESS. The rectangular upstream (left) vessel accommodates the electronics, cooling and pumping systems of SPICE. The L-shaped copper cold finger can be seen in orange. The beam enters from the left.
\label{fig:spice} 
\label{fig:coldfinger}
}
\end{figure*}

Internal conversion is a nuclear de-excitation process where the excitation energy of the nucleus is passed to an atomic electron, which is subsequently ejected from the atom. The energy of the ejected internal conversion electron (ICE) is equal to the nuclear transition energy minus the electron's binding energy. The binding energy is characteristic of the element and the atomic shell from which the electron was ejected. 
The rates of internal conversion for each atomic shell are highly sensitive to the multipolarity of the nuclear transition, with the relative rates being calculable from atomic theory (independent of nuclear structure effects). Hence, ICE spectrometers offer an alternative means to determine observables when other methods, such as high-statistics $\gamma$-ray angular distribution measurements, are not possible, or are insensitive to the physics of interest.

Electric monopole ($E0$) transitions, taking place between states of equal spin and parity, act as a probe 
of the degree of mixing and the difference in mean square charge radii ($\langle r^2 \rangle$) of nuclear configurations \cite{Wood1999}. $E0$ strength, particularly between low-lying $0^+$ states, is regarded as a key indicator of the shape-coexistence phenomenon widely occurring across the nuclear landscape \cite{heyde_rmp83}. 
$E0$ transitions via single-photon emission are strictly forbidden due to angular momentum conservation. Subsequently, electron-spectrometers are essential in the study of these transitions.
In view of the physics opportunities offered by electron spectroscopy, a number of devices with various designs have been constructed at facilities around the world \cite{kleinheinz_nim32,dionisio_nima437,luontama_nim159,Kibedi1990,butler_nima381,Pakarinen2014,vanKlinken_nim98,metlay_nima336,aengenvoort_epja1,Perkowski2014,Battaglia2016,Papadakis2018}.

The SPectrometer for Internal Conversion Electrons (SPICE) has been constructed and commissioned for use in conjunction with the TIGRESS spectrometer to perform combined in-beam $\gamma$-ray and ICE spectroscopy at TRIUMF's ISAC-II facility with stable and radioactive ion beams.
The basic concepts of the device were explored in detailed GEANT4 simulations by Ketelhut {\it et al.} \cite{ketelhut_nim214}. 
In this article, we describe the various components of the SPICE detector (Section \ref{sec:components}) and provide examples of the performance under typical operating conditions (Section \ref{sec:performance}).

\begin{figure}[!ht]
\centering
\fbox{\includegraphics[width=0.7\linewidth]{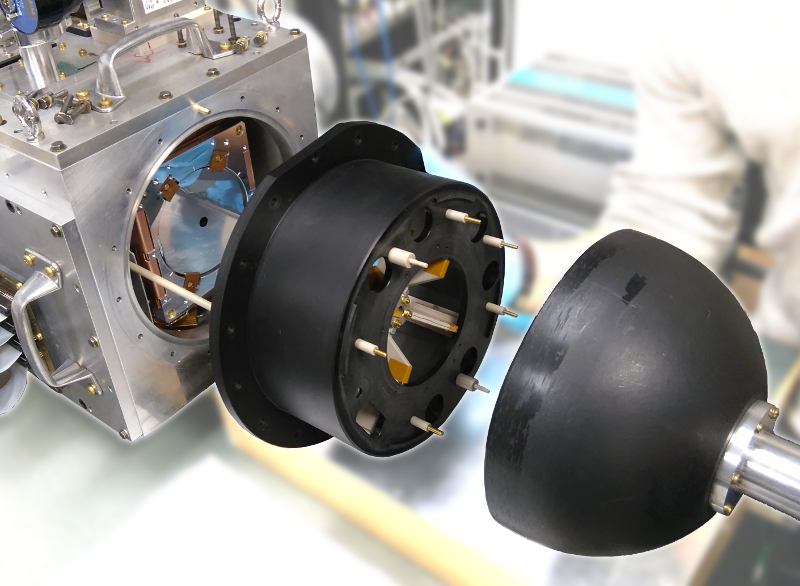}}
\caption{Photograph showing the aluminium vacuum vessel of SPICE, which houses the Si(Li) detector, and the two-part polymer target chamber containing the photon shield and magnet lens assembly. The target wheel (not pictured) sits inside the downstream dome portion.
\label{fig:spice2}}
\end{figure}

\section{Spectrometer Components}
\label{sec:components}

SPICE is an in-beam electron spectrometer, consisting of a single large magnetic lens, formed of rare-earth permanent magnets, and a highly segmented large-area lithium-drifted silicon [Si(Li)] detector located upstream of the reaction target. 
Fig.~\ref{fig:spice} shows SPICE coupled with the TIGRESS HPGe array.
The electron detector is shielded from direct view of the target by a photon shield of high-$Z$ material, which effectively suppresses beam-induced background radiation. The magnetic lens collects and guides electrons around the photon shield to the Si(Li) detector. This arrangement offers high detection efficiency over a wide range of electron energies, from $\approx$100 to $\approx$2000\,keV.

\begin{figure}[ht]
\centering
 \includegraphics[width=0.7\linewidth]{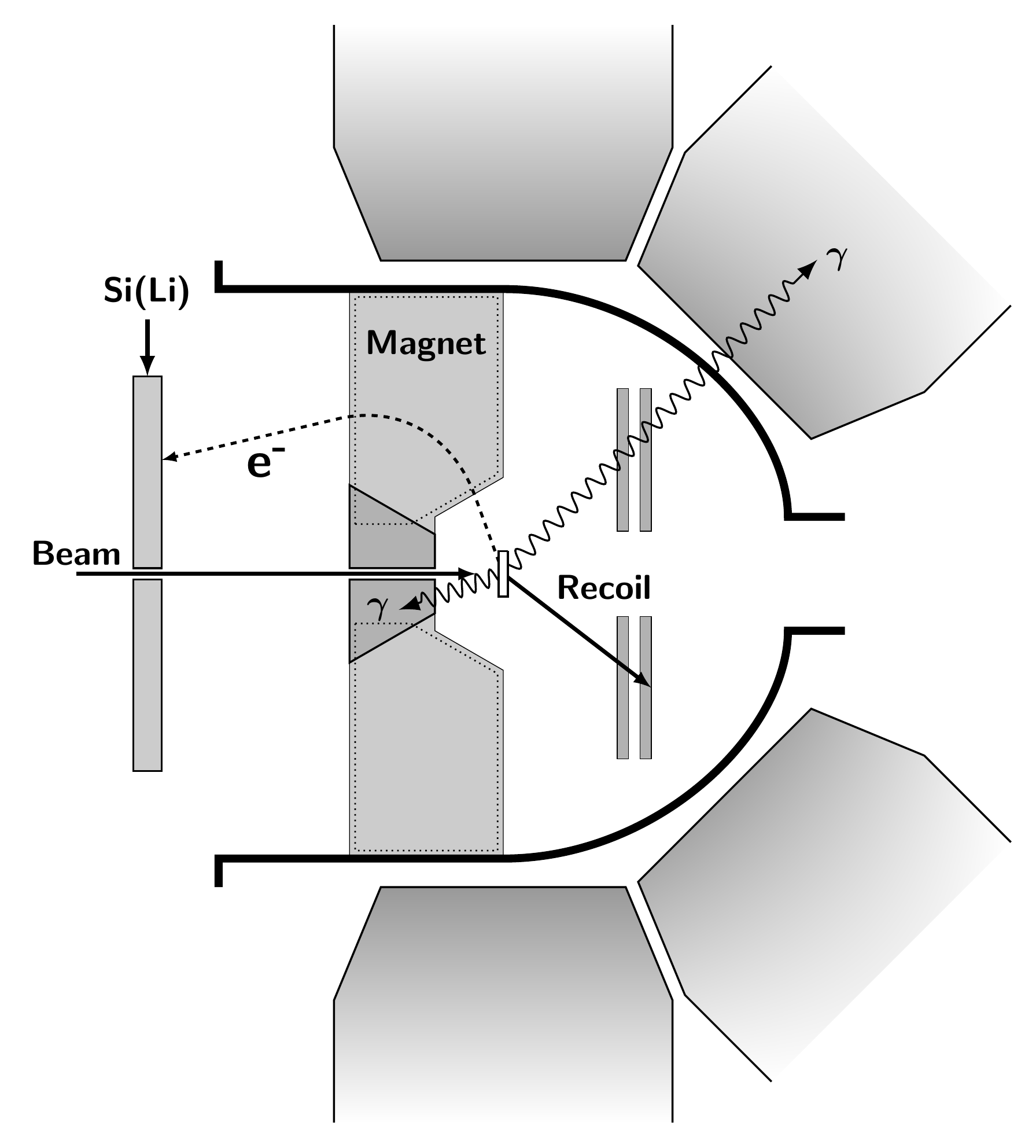}
\caption{Schematic diagram of the main elements of SPICE inside TIGRESS. The silhouettes of the central NdFeB magnets inside their plastic shrouds are shown by dotted lines. The target wheel and recoil detector are held in place by a PEEK (Polyether ether ketone) support structure which is not shown.
\label{fig:spiceschematic}}
\end{figure}

The components of SPICE are housed within three detachable volumes, pictured in Fig.~\ref{fig:spice2}. The Si(Li) detector is located inside a rectangular aluminium vessel which connects to the support structure and provides the cooling, pumping and electronics of SPICE. A liquid-nitrogen dewar is located on the top of the assembly, as seen in Fig.~\ref{fig:spice}.
The vacuum chamber is connected downstream to a polymer (Acetal copolymer) cylinder housing the magnetic lens and photon shield assembly.
The final volume consists of a dome-shaped polymer vacuum chamber, which facilitates access to the recoil detector and the targets mounted on the downstream face of the cylindrical chamber. Fig.~\ref{fig:spiceschematic} illustrates the relative positions of the spectrometer elements.
These two polymer volumes together form the reaction chamber of SPICE, which is surrounded by twelve Compton-suppressed TIGRESS clover detectors, four at 45$^\circ$ and eight at 90$^\circ$ with respect to the beam axis.

\subsection{Segmented Si(Li) Detector}
\begin{figure}[ht]
\fbox{\includegraphics[width=0.7\linewidth]{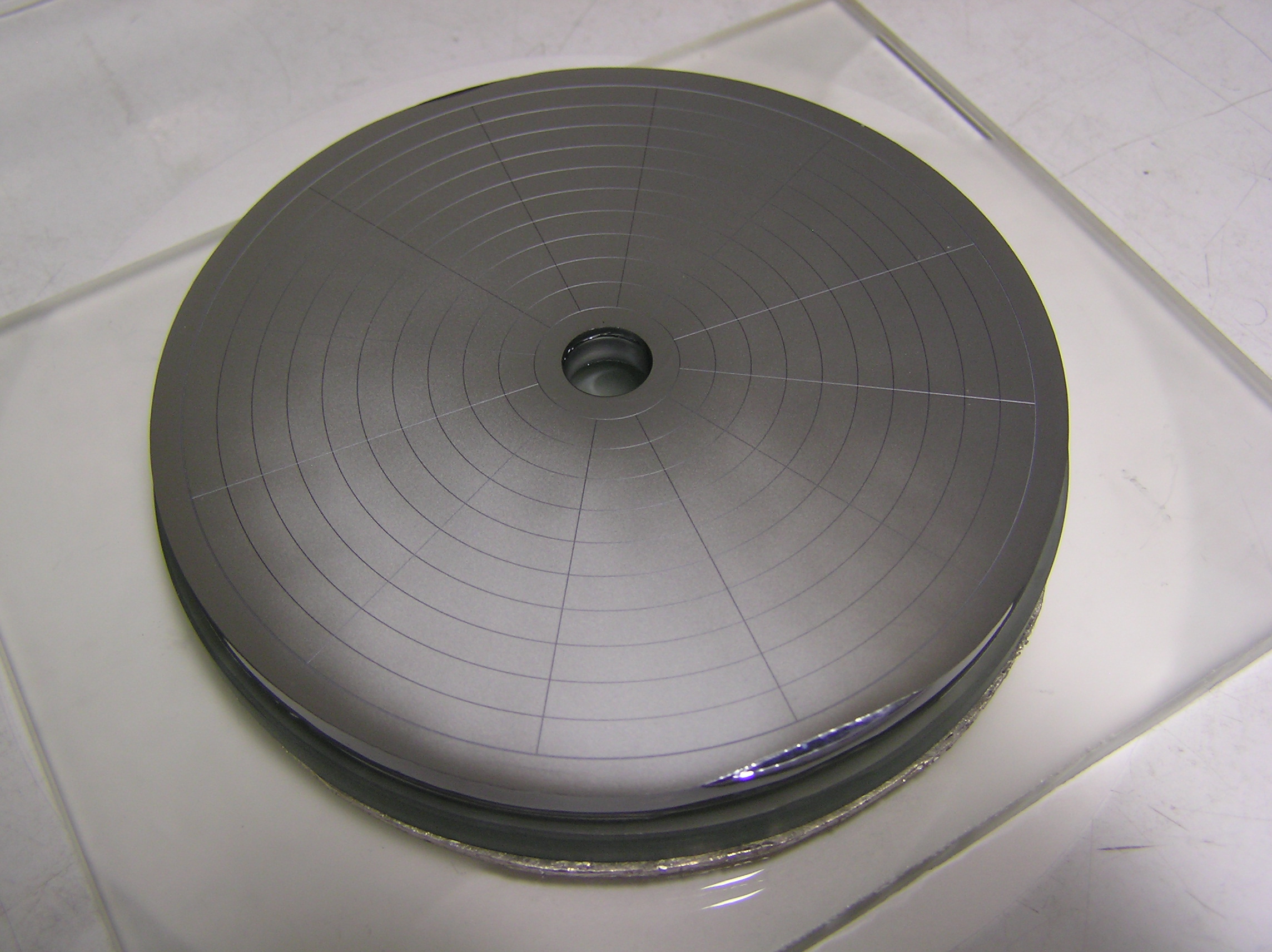}}
\centering
 \\
\caption{Photograph of the rear surface of the Si(Li) crystal. The laser-etched segmentation of the lithium-diffused (ohmic) contact can be seen. Image courtesy of SEMIKON Detector GmbH.}
\label{fig:SiLi}
\end{figure}

The internal conversion electrons are detected with an annular Si(Li) detector\footnote{The Si(Li) detector was custom designed and built by SEMIKON Detector GmbH (Karl-Heinz-Beckurts-Str. 13, 52428 J\"{u}lich, Germany; e-mail: info@semikondetector.de).}.
The detector has an active area of $\sim$67.4~cm$^2$, which is electronically segmented on the lithium-diffused (ohmic) contact by laser etching into 120 individual elements. The elements are arranged into ten equal rings of 3.9-mm pitch and twelve sectors of 30$^\circ$ pitch. The segmentation can be seen in Fig.~\ref{fig:SiLi}, a photograph taken before the silicon diode was installed into the detector housing.
Two guard rings surround the inner and outer perimeters of the active area to inhibit the circulation of surface leakage currents, with radial thicknesses of 3~mm and 4~mm, respectively. A central hole of 10-mm diameter allows passage of the beam.
The silicon wafer is $\approx$6.1~mm thick, which is sufficient for the detection of electron energies up to 3~MeV without primary electron punch-through. This has the added effect of significantly improving peak-to-total ratios for electrons of lower energy.
In normal operation, the junction contact is negatively biased at $-900$~V, applied in a DC-coupled configuration for noise suppression.

The detector is located 119~mm upstream of the central focal point of TIGRESS and the nominal target position, in the ``shadow" of the photon shield.
The detector is permanently mounted in an aluminium holder which encloses the signal-carrying printed circuit board (PCB). Detector segments are linked to the PCB by individual wire-bond connections.
The holder is mechanically supported by a cold-finger plate.
Two copper plates form the L-shaped cold-finger, shown in Fig.~\ref{fig:spice}. A large horizontal plate, with additional getter material fins, acts as both a heat conductor and as a cryogenic pump, supplementing a turbo-pump mounted on the side of the chamber.
The detector operates at a typical temperature of $-100^\circ$~C, monitored by a PT100 temperature sensor mounted in the detector holder.
At this temperature the leakage current is $<$1~nA, leading to an energy resolution of $\approx$4-keV FWHM at 1~MeV.
The vertical cold-finger plate is sandwiched by the detector holder and a second PCB on which 120 (one for each channel) field-effect transistors (JFETs) form the first stage of the preamplifier.
A tantalum collimator with an inner diameter of 8 mm is mounted on the cold finger upstream of the detector, this shields the detector from direct irradiation by the beam.

\subsection{Magnet Lens and Photon Shield}
\begin{figure}[ht]
 \centering
\fbox{\includegraphics[width=.5\linewidth]{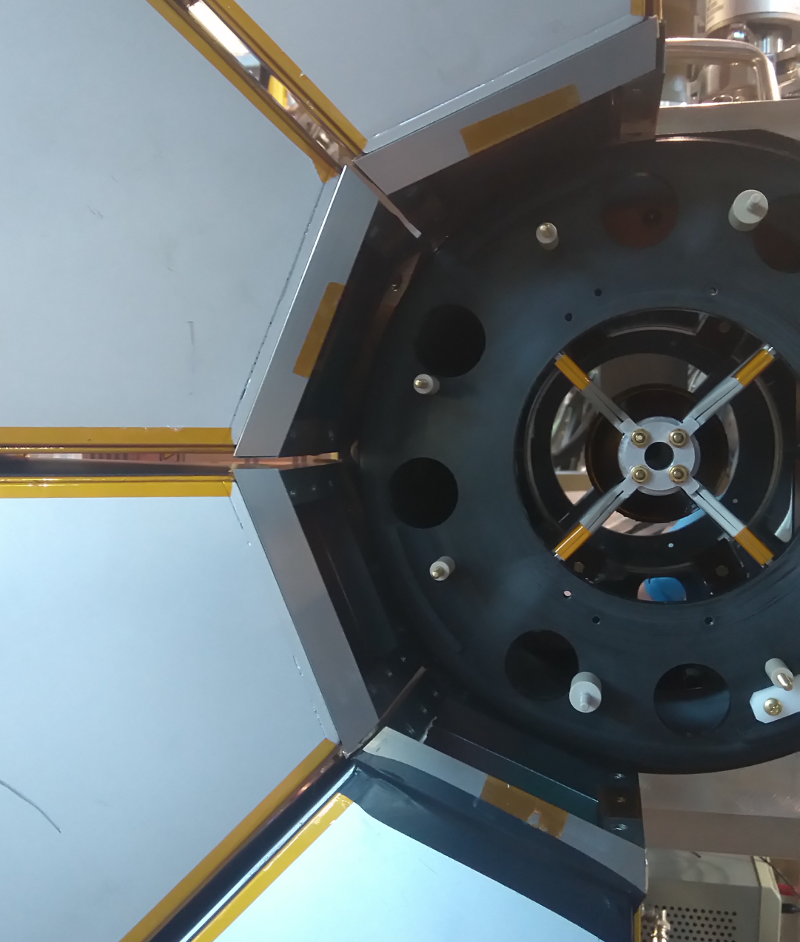}}
\centering
  \\
\caption{Photograph showing the upstream view of the magnet lens assembly. This image clearly shows the alignment of the lens with respect to the TIGRESS Compton-suppression shields for minimum impact on $\gamma$-ray efficiency.
\label{fig:MagnetLens}}
\end{figure}

The neodymium iron boron (NdFeB) permanent magnetic lens consists of four identical magnet clusters.
Each cluster consists of a central magnet in the shape of a rectangle cut at one corner (a three-right-angle pentagon) braced between a pair of shorter rectangular magnets to bolster the magnetic field towards the outer radius of the chamber.
This geometry can be seen in Fig.~\ref{fig:spiceschematic} and details can be found in Ref.~\cite{ketelhut_nim214}.
The four clusters are positioned at 90$^\circ$ intervals around the beam axis, as shown in Fig.~\ref{fig:MagnetLens}. The magnets are also positioned such that they align with the Compton-suppression shields of the 90$^\circ$ TIGRESS HPGe clovers. This arrangement allows for a negligible shadowing of the HPGe clovers by the magnet material, and results in a minimal impact on the $\gamma$-ray detection efficiency or spectral quality. 
The magnetization is arranged such that the lens assembly produces a clockwise inhomogeneous toroidal field when looking downstream.

The magnet clusters are enclosed inside 2-mm-thick PEEK covers to reduce the low-energy electron background induced by charged particles scattering in the high-$Z$ material of the neodymium. The sheathed magnet clusters are held in place at the wall of the polymer chamber and in slots of the photon shield which acts as the ``keystone" of the lens assembly.

The photon shield is a truncated cone, tapered at 22$^\circ$, with the front face located 28~mm from the nominal target position. A 6-mm-diameter central hole allows the beam to pass through the centre.
The shield is composed of three layers: a 25-mm-thick layer of a 97.5\% tantalum and 2.5\% tungsten alloy, followed by 4~mm of tin and 1~mm of copper.
The thickness of the Ta/W layer is sufficient to stop 99\% of incident 511-keV photons. The tin and copper layers absorb K X-rays produced from the Ta/W (65, 67~keV) and Sn (28~keV) layers, respectively. Additionally, the surface of the photon shield is coated with a 0.2-mm-thick layer of Kapton to reduce back-scattering of incident low-energy electrons. 

Small changes to the magnet specifications can be made to construct a lens with maximal efficiency in a region of interest for a specific experiment.
Two complete magnetic lens assemblies have been constructed, the low-energy lens (LEL) with a maximum field strength of 0.15~T and the medium-energy lens (MEL) with a maximum of 0.22~T. The MEL features thinner and longer rectangular lateral magnets, which significantly increases the strength of the magnetic field towards the inner radius of the chamber making the spectrometer suitable for the detection of higher-energy electrons up to $\sim$1500~keV.

\subsection{Target Wheel}
The target wheel consists of a rotatable aluminium disk mounted on a PEEK plate, which serves as a support structure for the target mechanism and recoil detectors.
It is mounted on stand-offs from the magnet lens chamber to ensure alignment between the two assemblies.
The target wheel holds two or more reaction targets and two tuning apertures.
The wheel can be rotated, by means of a gearing system connected to a mechanical vacuum-feedthrough located on the side of the upstream aluminium chamber, to select each reaction target or aperture without the need of breaking the vacuum, or warming-up the Si(Li) detector.
Where possible, low-$Z$ materials are used for all components to minimise $\gamma$-ray attenuation and the scattering of electrons.
Target foils are mounted on 0.5-mm-thick aluminium frames with an inner diameter of 8~mm.
Up to five targets can be mounted at the central focus of the HPGe detectors.
Alternately, up to two targets can be mounted on extended pedestals $8$\,mm upstream (closer to the magnet-lens).
In the latter case, a concentric $18$-mm-diameter hole in the target wheel allows transmission of recoiling heavy ions to the recoil detection system located in the downstream side of the chamber.
The upstream target position enhances the detection efficiency of electrons emitted in-flight from excited states with lifetimes in the range of several picoseconds to nanoseconds. In this case, the decay takes place in-flight between the reaction target and the target wheel plane. For longer lifetimes, a thick target can be used or a catcher foil can be mounted at the target wheel to stop the recoiling nuclei and detect their decay at rest.

The target wheel was originally designed with the capability of applying a bias voltage to each target frame for the purpose of suppressing $\delta$-electron production. However, this feature has not been used because the design of the magnetic lens and photon shield alone has provided sufficient suppression.

\subsection{Signal Shaping and Amplification}
SPICE uses a custom-designed two-stage charge sensitive amplifier.
The cooled front-end stage (located close to the  Si(Li) detector on the common cold finger) is composed of 120 JFETs (type BF862), operated in a common source configuration\footnote{JFETs use gate feedback components R$_{f}$=30~M$\Omega$ and C$_{f}$=1.2~pF}.
Outputs from the JFETs are grouped into 8 sets of 15 signal-feedback\footnote{The second-stage amplifier provides feedback to discharge the gate of the JFET.} pairs.
Unsheathed coaxial cables connect to the second-stage amplifiers on 8 external cards, located outside the vacuum on the top side of the aluminium chamber.
A third-stage buffer amplifier located on the cards provides gain and drives the output over 50~$\Omega$ coaxial cables to the data acquisition system.
Typical output signals have a risetime of $\approx$200~ns and a decay constant of $\approx$40~$\mu$s. The gain is $\approx$200~mV/MeV.

Output signals from the preamplifiers are processed by TIG-10 100~MHz, 14-bit digitizer cards of the TIGRESS data acquisition system \cite{martin_ieee55}. Real-time signal processing provides energy and timing information for each event.
The option of collecting waveforms is available for more sophisticated offline processing.

\begin{figure*}[htb]
\centering
 \includegraphics[width=\linewidth]{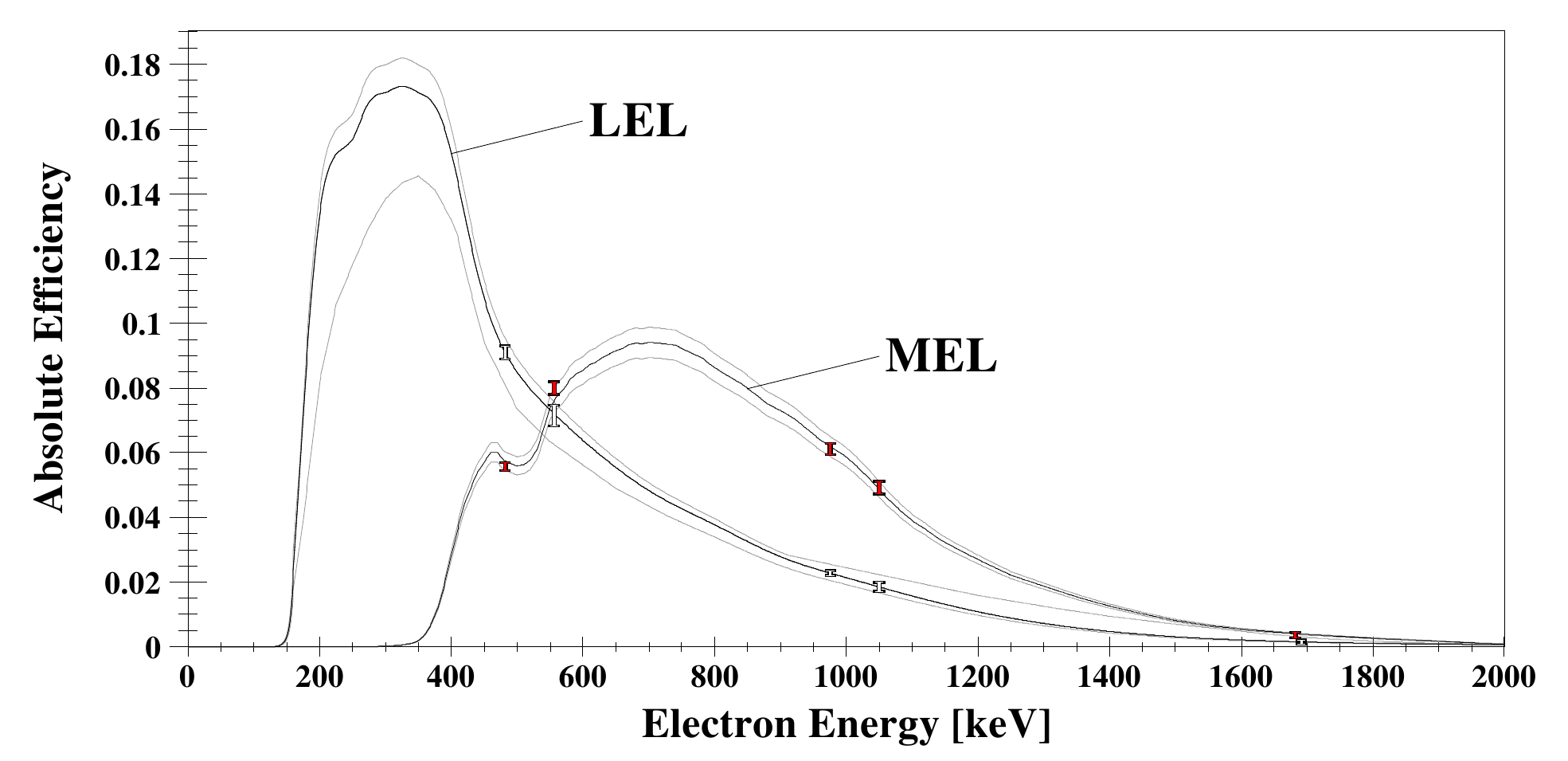}
\caption{Absolute efficiency of SPICE with the low-energy and the medium-energy magnetic lens configurations.
GEANT4 simulated efficiency curves are fit to experimental data points of transitions from a radioactive $^{207}$Bi conversion electron source.
The grey lines indicate the typical variation in the absolute efficiency which can result from different experimental conditions, determined from detailed simulations of these effects (see text for details).
\label{fig:Eff}}
\end{figure*}

\subsection{Downstream Recoil Detectors}
The downstream dome of the SPICE target chamber has an inner radius of 97~mm to accommodate recoil charged-particle detectors.
SPICE is typically operated with a Micron-Semiconductor S3-type annular double-sided silicon-strip detector \cite{Micron}.
Nominally positioned 2.9~cm downstream of the target, an S3 has an angular coverage of 21-50$^\circ$.
Detectors are mechanically supported on the target wheel by three brass rods and PEEK spacers.
Shorter or longer spacers can be used to adjust the angular range up to a maximum recoil angle of $\approx$60$^\circ$ before there is interference with the target wheel.
The ohmic contact of an S3 is divided into 32 azimuthal sectors and the junction contact into 24 rings providing a polar angular resolution of $\leq$1.7$^\circ$ when placed at the nominal distance.
Signals are carried by a 64-channel high-density ribbon cable routed along the outer edge of the inside of the magnet chamber enclosed in a low-profile plastic conduit.
On a side panel of the aluminium vacuum vessel, the signals are separated into four 16-channel pre-amplifiers purchased from Swan Research \cite{SWAN}.
Two sets of preamplifiers are available with suitable gain for the detection of either low-energy charged particles or high-energy heavy ion recoils. 
An additional S3-type detector can be mounted immediately downstream\footnote{The second S3 is mounted on the same brass rods, 1~mm plastic spacers between the two PCBs ensure isolation of the signal tracks. Both detector signal cables share the same conduit.} of the first to perform $\Delta E-E$ identification of light charged particles, in order to provide reaction-channel selectivity. Such a configuration has been operated using 150~$\mu$m thickness $\Delta E$ and 1000~$\mu$m $E$ detectors for identification of protons, deuterons, tritons and $\alpha$-particles at energies of 5 to 40~MeV (See Section \ref{sec:perform:coinc}).

In the future, other ancillary detector systems, which have been designed to couple with the TIGRESS spectrometer on the downstream side can be combined with the SPICE+TIGRESS setup. Examples include the wall of CsI(Tl) detectors described in Ref.~\cite{voss_nim214} which offers particle identification with radiation hardness, the DESCANT \cite{Bildstein2015} neutron array and the recently commissioned EMMA \cite{davids_nim205} electromagnetic separator, which would enable additional sensitivity in channel selection.

\begin{figure}[!t]
\centering
 \includegraphics[width=0.9\linewidth]{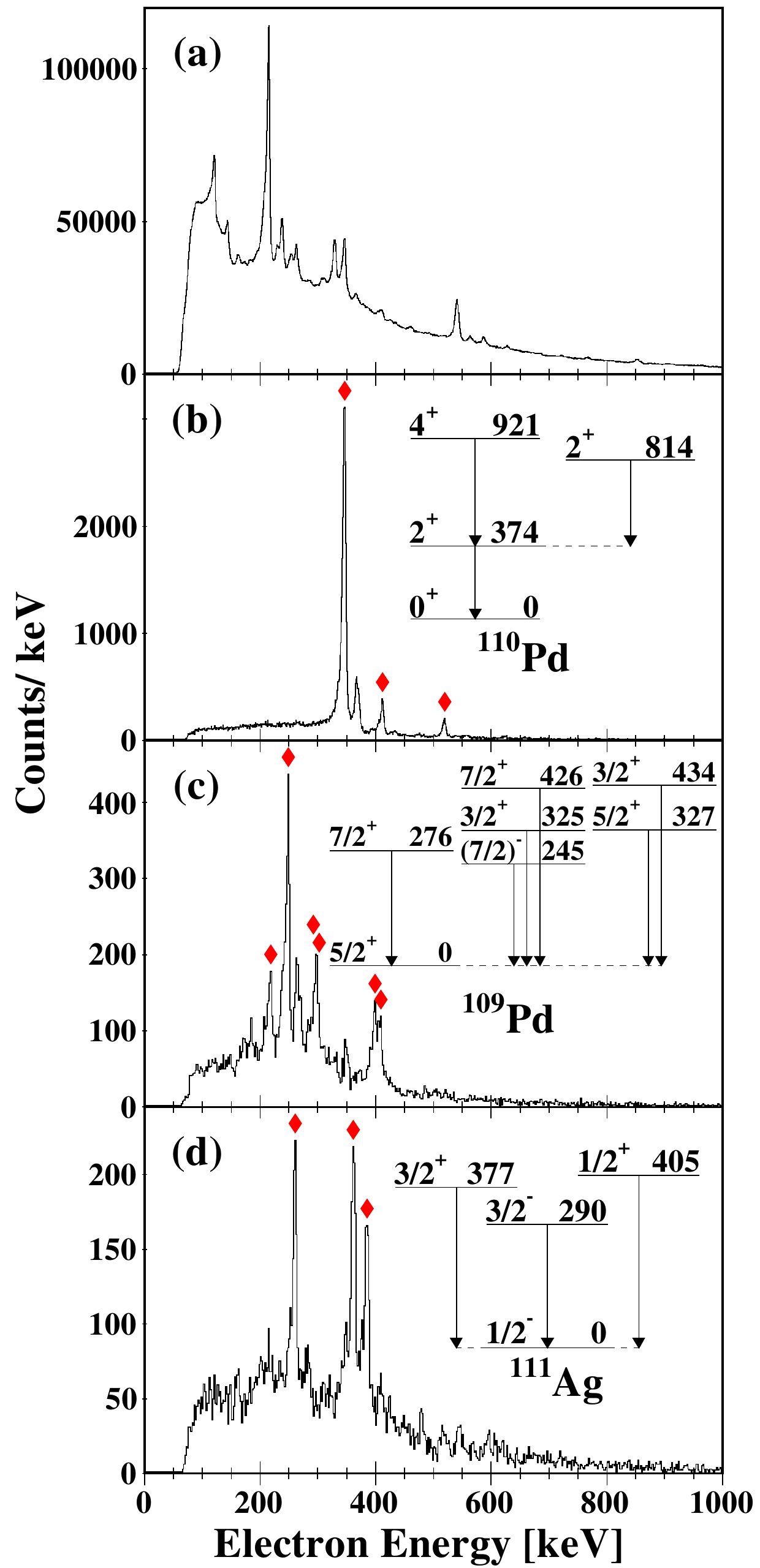}
\caption{ICE spectra from the $^{110}$Pd+$\alpha$ reaction demonstrating the degree to which ICE from a single channel can be isolated with particle gating.
(a) The ungated ICE spectrum, showing good peak-to-total ratio but high reaction-channel density.
(b) ICE spectrum in coincidence with inelastically scattered $\alpha$ particles detected in the downstream charged-particle telescope, highlighting transitions in $^{110}$Pd.
(c) ICE spectrum in coincidence with low-energy $\alpha$ particles, an energy cut preferentially selects the $Q$-value of the $^{110}$Pd($\alpha$,$\alpha$+n)$^{109}$Pd channel.
(d) ICE spectrum in coincidence with a detected triton, selecting the $^{110}$Pd($\alpha$,t)$^{111}$Ag channel.
Partial level schemes highlight visible ICE transitions. Markers indicate the corresponding $K$ electrons peaks.
\label{fig:coinc}}
\end{figure}

\section{Spectrometer Performance}
\label{sec:performance}
The performance of SPICE has been characterised both with sources and during in-beam measurements conducted at TRIUMF.
Two experiments have been conducted to demonstrate the capabilities of SPICE.
A beam of 36-MeV $\alpha$ particles delivered by the TRIUMF-ISAC-II superconducting linear accelerator \cite{Laxdal2014} was used to induce multiple reactions in a self-supporting $^{110}$Pd foil target (1.6\,mg/cm$^2$, 99.0\% isotopic purity). In this beam time, the LEL was used and two S3 detectors were placed downstream in the $\Delta E-E$ telescope configuration to identify light ejectiles.
In a separate beam time, $^{12}$C ions at 68\,MeV were incident upon targets of $^{152}$Sm (4\,mg/cm$^2$, $>$99\% isotopic purity) and $^{196}$Pt (2.93\,mg/cm$^2$, 94.6\% isotopic purity). The LEL was used and a single S3 detector was located downstream to detect $^{12}$C beam recoils. Results from these tests are given as examples below to highlight key features and capabilities of SPICE.

\subsection{Electron-detection Efficiency}\label{sec:efftalk}
In each experiment, one of the permanent magnetic lens assemblies must be selected such that the SPICE detection efficiency is best suited for the energy range of interest.
The transport efficiency of both of the presently available magnetic lenses falls sharply for low-energy electrons. This feature is intentional and ensures strong suppression of the $\delta$-electron background produced from the target during in-beam measurements. Low-energy electrons are directed into the photon shield where they are stopped and never reach the Si(Li) detector.

\begin{figure*}[htb]
\centering
 \includegraphics[width=\linewidth]{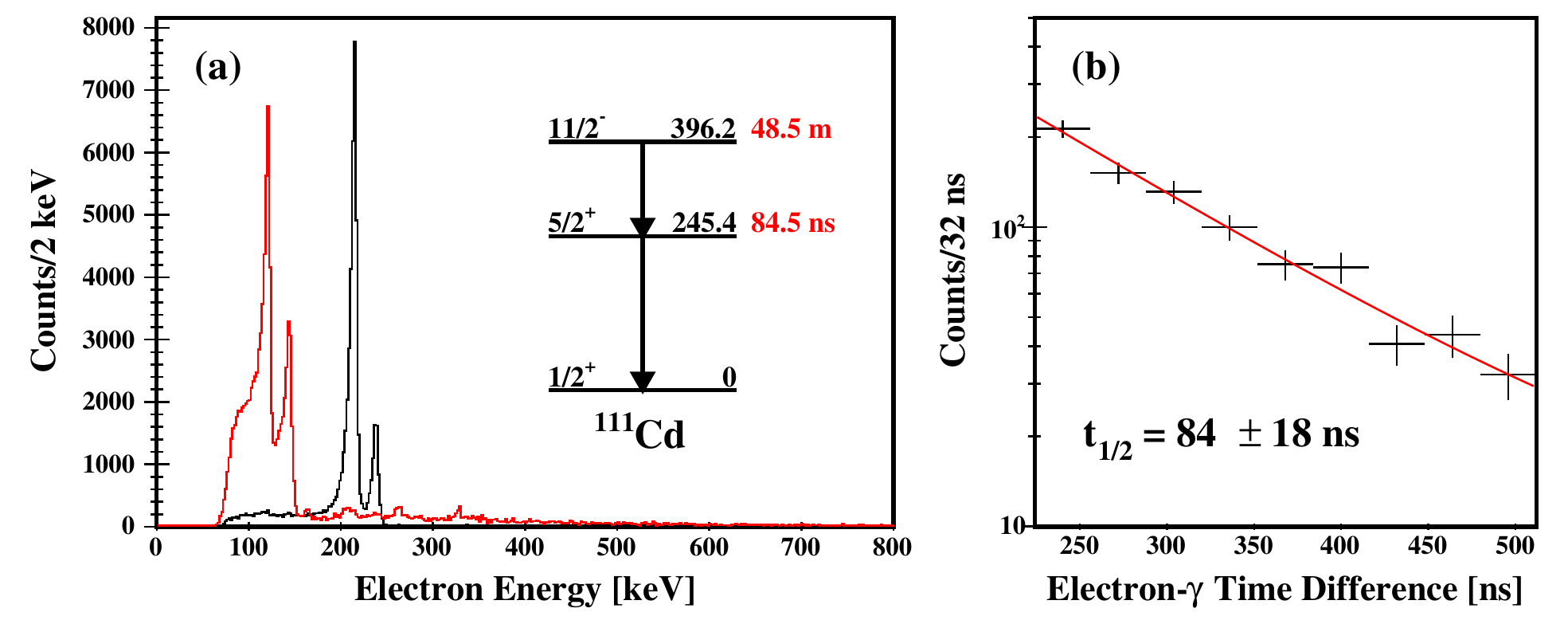}
\caption{(a) Coincidence $\gamma$-ray gated ICE energy spectra from the decay of isomeric states in $^{111}$Cd populated by the $^{110}$Pd($\alpha$,3n)$^{111}$Cd reaction. The ICE spectrum resulting from the 245.4\,keV $\gamma$-ray gate (red) has some additional structure from prompt feeding, whereas the spectra gated on the 150.8\,keV $\gamma$ ray from the 48.5-minute isomer (black) is extremely clean.
(b) Time difference between a 150.8\,keV $\gamma$ ray and a $K$ electron from the 245.4\,keV transition, giving a measurement of the 5/2$^+$ state half-life.
\label{fig:cdisomer}}
\end{figure*}

The detection efficiency for internal conversion electrons using both the low- and medium-energy magnetic lenses has been carefully studied with a series of detailed COMSOL and GEANT4 simulations \cite{ketelhut_nim214,comsol_2012,ALLISON2016186}. The COMSOL calculations of the magnetic field strength were optimised to reproduce precise Hall-probe mapping of the inhomogeneous field.
Figure~\ref{fig:Eff} shows the simulated efficiency compared to absolute efficiency measurements using a NIST calibrated $^{207}$Bi source. The simulation results shown here include the specific geometry of the calibration source and holder.
The peak efficiency of the MEL, at 700~keV, is almost 10\%, and the efficiency at 1~MeV is $>$5\%. The good efficiency at higher electron energies is crucial for studies in $A$$<$100 nuclei. The efficiency drops to 1\% by 1500~keV, however, at higher energies the dominance of internal pair formation usually makes ICE spectroscopy untenable. The efficiency of the LEL peaks at 350~keV and is greater than 14\%.

Due to the strong and complex magnetic field of the target region, small differences in beam-spot position and size and target frames can modify the efficiency. These effects are particularly notable when using the low-energy lens. The grey lines in Figure~\ref{fig:Eff} indicate the typical variation in the absolute efficiency which has been found from comparing a series of simulation results for different experimental conditions. The simulations explore the effects of reasonable differences in target or beam-spot position, as well as beam-spot diameter and reaction kinematics. It is found that both the exact shape of the curve as well as the absolute scaling can change between different experimental conditions and source measurements. There may also be additional energy-dependent electronic effects which can have an impact on the efficiency. As such, while simulations of the specific experimental conditions can provide the shape and magnitude of the efficiency curve, it is still important that any results from simulations be normalized to known transitions observed in the in-beam data for the energy region of interest\footnote{Any fits should not be extrapolated beyond available data due to rapid changes in the gradient of the efficiency curve.}.
This normalisation is typically achieved by using stretched $E2$ transitions in the yrast-band of a well-known even-even nucleus.

Using the well-understood TIGRESS efficiency, $\epsilon_\gamma$, at a particular $\gamma$-ray energy and the theoretical ICC from BrICC, $\alpha(E2)_{BrICC}$, the in-beam SPICE efficiency, $\epsilon_{e^-}$, can be deduced for normalisation using:
\begin{equation}
\alpha(E2)_{BrICC}=
\dfrac{N_{e^-}\epsilon_\gamma}{N_\gamma\epsilon_{e^-}},
\end{equation}
\noindent where $N_{e^-}$ and $N_{\gamma}$ are the number of counts detected for the electrons and $\gamma$ rays, respectively. If it cannot be obtained from the main experimental data, then sufficient data for such a normalisation can typically be accumulated in just a few hours using a stable beam at the beginning or end of a RIB experiment.

\subsection{Coincidence Measurements}
\label{sec:perform:coinc}

SPICE is intended to utilise coincidence measurements to increase the peak-to-total spectral quality and identify weak channels. This method works in unison with other features, such as the photon shield, which is designed to reduce background.

The recoil chamber of SPICE is suitable for a range of charged-particle detectors, which can cover scattering angles up to 60$^\circ$ for performing electron-ion coincidence measurements.
Figure \ref{fig:coinc} shows the power of this technique with data collected during the $\alpha$+$^{110}$Pd beam time.
A charged-particle telescope of two S3 detectors was used to select the reaction product by identification of various charged particles; $^{110}$Pd($\alpha$,$\alpha'$)$^{110}$Pd*, $^{110}$Pd($\alpha$,$\alpha$+n)$^{109}$Pd* and $^{110}$Pd($\alpha$,t)$^{111}$Ag*. Each of these reaction products are highlighted in panels (b)-(d) of Figure \ref{fig:coinc}, respectively. The latter two examples show the quality of the electron spectra which can be obtained even for weak reaction channels.

\begin{figure}[t]
\centering
 \includegraphics[width=0.98\linewidth]{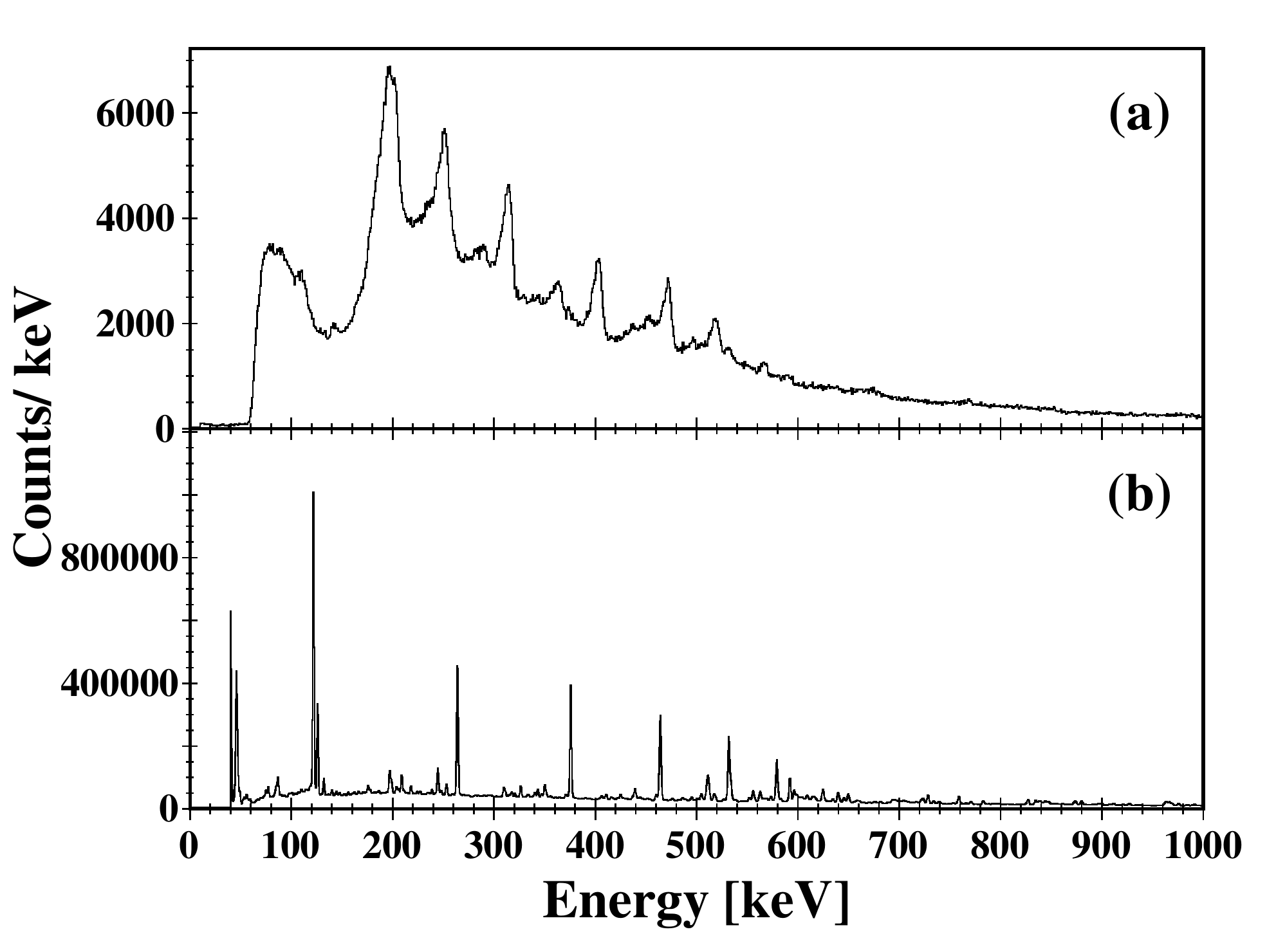}
\caption{SPICE (a) and TIGRESS (b) singles energy spectra from inelastic scattering and fusion-evaporation reactions induced by a $^{12}$C beam incident on a $^{152}$Sm target. The spectra are dominated by the $E2$ transitions of the $^{160}$Er (4n channel) yrast band. Inelastic excitation of the $2_1^+$ state of $^{152}$Sm can also be seen.
\label{fig:160Er}}
\end{figure}

Alternatively to, or in combination with, charged-particle coincidence gating, further selectivity can be achieved by requiring a specific $\gamma$-ray detected in TIGRESS.
SPICE is coupled to 12 HPGe TIGRESS clovers with a $\gamma$-ray detection efficiency of $\approx$9\% at 1~MeV when operating in high-efficiency mode.
In the same $\alpha$+$^{110}$Pd reaction described above, coincidence $\gamma$-ray gating was able to select electrons from neutron-only channels $^{110}$Pd($\alpha$,3n)$^{111}$Cd and $^{110}$Pd($\alpha$,4n)$^{110}$Cd.
Fig.~\ref{fig:cdisomer} shows the $\gamma$-ray gated ICE spectra for the decay of isomeric states in $^{111}$Cd populated in the reaction.
The ungated ICE spectrum is shown in Figure \ref{fig:coinc}a.
From these extremely clean $\gamma$-ray gated ICE spectra, it is possible to perform ICE-$\gamma$ timing coincidence measurements of long-lived states. For the 245.4-keV 5/2$^+$ state in $^{111}$Cd, a half-life of 84$\pm$18~ns was extracted in agreement with the literature value of 84.5$\pm$0.4~ns.
Such techniques might be used, for example, in the case of a long lived $0^+\rightarrow0^+$ decay, to simultaneously extract both a lifetime and an electron branching ratio in order to determine the $E0$ transition strength.

\subsection{Resolving Power}
\begin{figure}[t]
\centering
 \includegraphics[width=0.98\linewidth]{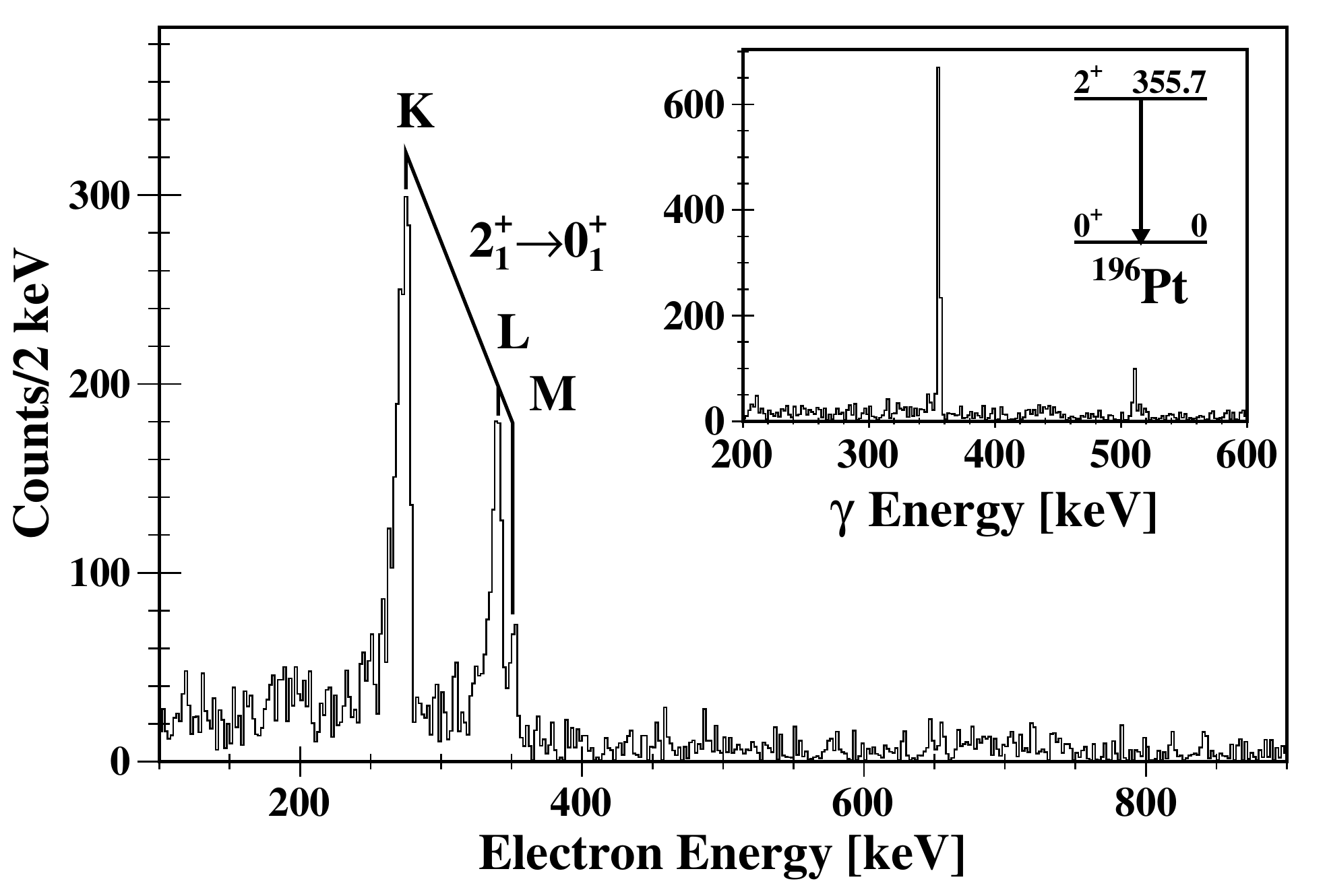}
\caption{SPICE spectrum from Coulomb excitation of $^{196}$Pt gated on $^{12}$C recoils.
The $K$- $L$- and $M$-shell peaks of the $2^+_1 \rightarrow 0^+_1$ transition in $^{196}$Pt are clearly visible.
The inset image shows the equivalent $\gamma$-ray spectrum from TIGRESS.
\label{fig:196Pt}}
\end{figure}

The in-beam resolving power of SPICE was demonstrated during the $^{12}$C+$^{196}$Pt Coulomb excitation study.
The reaction was dominated by a single-step excitation of the $2_1^+$ 355.7-keV state in $^{196}$Pt.
De-excitation $\gamma$ rays and ICEs are shown in Figure \ref{fig:196Pt}, where a gate has been placed on detected $^{12}$C recoils to remove background from other reaction channels.
The SPICE electron spectrum shows that the $K$-, $L$- and $M$-shell ICE lines for the $2^+_1 \rightarrow 0^+_1$ transition in $^{196}$Pt are clearly resolved.
With sufficient statistics, an in-beam calibration could improve upon this resolution by precisely aligning the energy response of all detector segments. Note that energy straggling in the thick, high-$Z$ target foil, adds a significant exponential tail to the ICE peaks, particularly at low energies. These tails must be accounted for in peak fitting. The peaks observed here have a FWHM of 6.7~keV at 300~keV, which is slightly worse than the $4.2$-keV FWHM at 1~MeV achieved with a $^{207}$Bi source.

The good energy resolution made it possible to perform a measurement of the conversion ratios of the different electron shells.
In-beam normalisation of the efficiency curve was achieved using the $^{152}$Sm target of the same beam time.
States up to the 4022-keV $18_1^+$ state in $^{160}$Er were populated by the $^{152}$Sm($^{12}$C,4n)$^{160}$Er* reaction and the subsequent $E2$ decays within the yrast band were observed in SPICE and TIGRESS, as shown in Figure \ref{fig:160Er}, providing normalisation points.
The resulting $K$-, $L$- and $M$-shell ratios for $^{196}$Pt are given in Table \ref{tab:196Pt}. The values measured with SPICE are in good agreement with the theoretical values calculated using BrICC \cite{Kibedi2008}.

\begin{table}[t]
\caption{Internal conversion measurements of the 355.7\,keV ${2^+_1\rightarrow0^+_1}$ transition following Coulomb excitation of $^{196}$Pt compared with calculated values from BrICC.} 
\centering 
\begin{tabular}{c|S[table-format=1.2]S[table-format=1.2]} 
\hline 
 & $\alpha_K$/$\alpha_L$ & $\alpha_L$/$\alpha_{M+N+O}$ \\ [0.5ex] 
\hline 
SPICE & 2.7\,(3) &  2.8\,(3)  \\
BrICC \cite{Kibedi2008} & 2.65\,(6) &3.13\,(5)   \\ 
\hline 
\end{tabular}
\label{tab:196Pt} 
\end{table}

\section{Summary}
\label{sec:summary}
The SPectrometer for Internal Conversion Electrons (SPICE) has been commissioned at TRIUMF's ISAC-II facility for use in conjunction with the TIGRESS $\gamma$-ray spectrometer. SPICE uses a permanent rare-earth magnetic lens to collect and direct internal conversion electrons emitted from the target position around a photon shield to a thick and highly segmented lithium-drifted silicon detector. This arrangement, combined with  TIGRESS, enables high-resolution in-beam $\gamma$-ray and internal conversion electron spectroscopy to be performed with stable and radioactive ion beams. Commissioning runs have demonstrated that the design of SPICE significantly suppresses the flux of $\delta$ electrons, which typically hinders in-beam electron measurements. The initial performance shows excellent channel selection, with high resolving power and good detection efficiency for electrons in the medium-energy range, which is ideal for studies of electric monopole transitions.

\section{Acknowledgements}
We would like to thank the beam delivery and technical staff of the TRIUMF-ISAC facility for providing the beams used in the development of SPICE. The infrastructure of SPICE has been funded by the Canada Foundation for Innovation and the Ontario Ministry of Research and Innovation, with personnel contributions from TRIUMF. TRIUMF is funded through a contribution agreement with the National Research Council of Canada. C.E.S. acknowledges support from the Canada Research Chairs program. This work was partially supported by the Natural Sciences and Engineering Research Council of Canada and by the National Science Foundation Grant No. 1606890.
The enriched isotope used in this research was supplied by the United States Department of Energy Office of Science by the Isotope Program in the Office of Nuclear Physics.


\Urlmuskip=0mu plus 1mu\relax
\bibliographystyle{elsarticle-num}
\bibliography{main}

\begin{thebibliography}{10}
\expandafter\ifx\csname url\endcsname\relax
  \def\url#1{\texttt{#1}}\fi
\expandafter\ifx\csname urlprefix\endcsname\relax\def\urlprefix{URL }\fi
\expandafter\ifx\csname href\endcsname\relax
  \def\href#1#2{#2} \def\path#1{#1}\fi

\bibitem{svensson_jpg31}
C.~E. Svensson, P.~Amaudruz, C.~Andreoiu, A.~Andreyev, R.~A.~E. Austin, G.~C.
  Ball, D.~Bandyopadhyay, A.~J. Boston, R.~S. Chakrawarthy, A.~A. Chen,
  R.~Churchman, T.~E. Drake, P.~Finlay, P.~E. Garrett, G.~F. Grinyer,
  G.~Hackman, B.~Hyland, B.~Jones, R.~Kanungo, R.~Maharaj, J.~P. Martin,
  D.~Morris, A.~C. Morton, C.~J. Pearson, A.~A. Phillips, J.~J. Ressler,
  R.~Roy, F.~Sarazin, M.~A. Schumaker, H.~C. Scraggs, M.~B. Smith,
  N.~Starinsky, J.~J. Valiente-Dob\'on, J.~C. Waddington, L.~M. Watters,
  {TIGRESS}: {TRIUMF}-{ISAC} $\gamma$-ray escape-suppressed spectrometer,
  Journal of Physics G: Nuclear and Particle Physics 31~(10) (2005) S1663.

\bibitem{hackman_hfi214}
G.~Hackman, C.~E. Svensson, The {TRIUMF}-{ISAC} gamma-ray escape suppressed
  spectrometer, {TIGRESS}, Hyperfine Interact. 225 (2014) 241.

\bibitem{davids_nim205}
B.~Davids, C.~N. Davids, {EMMA}: A recoil mass spectrometer for {ISAC-II} at
  {TRIUMF}, Nuclear Instruments and Methods 544 (2005) 565.

\bibitem{diget_jinst211}
C.~A. Diget, S.~P. Fox, A.~Smith, S.~Williams, M.~Porter-Peden, L.~Achouri,
  P.~Adsley, H.~Al-Falou, R.~A.~E. Austin, G.~C. Ball, J.~C. Blackmon,
  S.~Brown, W.~N. Catford, A.~A. Chen, J.~Chen, R.~M. Churchman, J.~Dech, D.~D.
  Valentino, M.~Djongolov, B.~R. Fulton, A.~Garnsworthy, G.~Hackman, U.~Hager,
  R.~Kshetri, L.~Kurchaninov, A.~M. Laird, J.~P. Martin, M.~Matos, J.~N. Orce,
  N.~A. Orr, C.~J. Pearson, C.~Ruiz, F.~Sarazin, S.~Sjue, D.~Smalley, C.~E.
  Svensson, M.~Taggart, E.~Tardiff, G.~L. Wilson, {SHARC}: Silicon
  highly-segmented array for reactions and coulex used in conjunction with the
  {TIGRESS} $\gamma$-ray spectrometer, JINST 6 (2011) P02005.

\bibitem{voss_nim214}
P.~Voss, R.~Henderson, C.~Andreoiu, R.~Ashley, R.~A.~E. Austin, G.~C. Ball,
  P.~C. Bender, A.~Bey, A.~Cheeseman, A.~Chester, D.~S. Cross, T.~E. Drake,
  A.~B. Garnsworthy, G.~Hackman, R.~Holland, S.~Ketelhut, P.~Kowalski, R.~Kr{\"
  u}cken, A.~T. Laffoley, K.~G. Leach, D.~Miller, W.~J. Mills, M.~Moukaddam,
  C.~J. Pearson, J.~Pore, E.~T. Rand, M.~M. Rajabali, U.~Rizwan, J.~Shoults,
  K.~Starosta, C.~E. Svensson, E.~Tardiff, C.~Unsworth, K.~V. Wieren, Z.-M.
  Wang, J.~Williams,
  \href{http://www.sciencedirect.com/science/article/pii/S0168900214001429}{The
  {TIGRESS} integrated plunger ancillary systems for electromagnetic transition
  rate studies at {TRIUMF}}, Nuclear Instruments and Methods 746 (2014) 87 --
  97.
\newblock \href {http://dx.doi.org/https://doi.org/10.1016/j.nima.2014.02.006}
  {\path{doi:https://doi.org/10.1016/j.nima.2014.02.006}}.
\newline\urlprefix\url{http://www.sciencedirect.com/science/article/pii/S0168900214001429}

\bibitem{Chester2018}
A.~Chester, G.~C. Ball, N.~Bernier, D.~S. Cross, T.~Domingo, T.~E. Drake, L.~J.
  Evitts, F.~H. Garcia, A.~B. Garnsworthy, G.~Hackman, S.~Hallam, J.~Henderson,
  R.~Henderson, R.~Kruecken, E.~MacConnachie, M.~Moukaddam, E.~Padilla-Rodal,
  O.~Paetkau, J.~L. Pore, U.~Rizwan, P.~Ruotsalainen, J.~Shoults,
  J.~Smallcombe, J.~K. Smith, K.~Starosta, C.~E. Svensson, K.~V. Wieren,
  J.~Williams, M.~Williams,
  \href{http://www.sciencedirect.com/science/article/pii/S0168900217312275}{Recoil
  distance method lifetime measurements at {TRIUMF}-{ISAC} using the {TIGRESS}
  integrated plunger}, Nuclear Instruments and Methods in Physics Research
  Section A: Accelerators, Spectrometers, Detectors and Associated Equipment
  882 (2018) 69 -- 83.
\newblock \href {http://dx.doi.org/https://doi.org/10.1016/j.nima.2017.11.029}
  {\path{doi:https://doi.org/10.1016/j.nima.2017.11.029}}.
\newline\urlprefix\url{http://www.sciencedirect.com/science/article/pii/S0168900217312275}

\bibitem{Wood1999}
J.~L. Wood, E.~F. Zganjar, C.~D. Coster, K.~Heyde, Electric monopole
  transitions from low energy excitations in nuclei, Nuclear Physics A 651
  (1999) 323 -- 368.

\bibitem{heyde_rmp83}
K.~Heyde, J.~L. Wood, Shape coexistence in atomic nuclei, Rev. Mod. Phys. 83
  (2011) 1467--1521.

\bibitem{kleinheinz_nim32}
P.~Kleinheinz, L.~Samuelsson, R.~Vukanović, K.~Siegbahn, A four-detector
  electron directional correlation spectrometer, Nuclear Instruments and
  Methods 32~(1) (1965) 1 -- 27.

\bibitem{dionisio_nima437}
J.~S. Dionisio, C.~Vieu, E.~Gueorguieva, M.~Kaci, E.~B. Kharraja, M.~G.
  Porquet, C.~Schück, J.~M. Lagrange, M.~Pautrat, W.~Phillips, J.~Durell,
  P.~G. Dagnall, S.~J. Dorning, M.~A. Jones, A.~G. Smith, B.~J. Varley,
  J.~C.~S. Bacelar, W.~Urban, T.~Rzaca-Urban, A.~Minkova, T.~Venkova,
  H.~Folger, J.~Vanhorenbeeck, A.~Passoja, Recent developments of multi e-γ
  spectrometers, Nuclear Instruments and Methods in Physics Research Section A:
  Accelerators, Spectrometers, Detectors and Associated Equipment 437~(2–3)
  (1999) 282 -- 334.

\bibitem{luontama_nim159}
M.~Luontama, J.~Kantele, R.~Julin, A.~Passoja, T.~Poikolainen,
  M.~Pylvänäinen, A combination intermediate-image magnetic plus {Si(Li)}
  electron spectrometer for in-beam experiments, Nuclear Instruments and
  Methods 159~(2–3) (1979) 339 -- 345.

\bibitem{Kibedi1990}
T.~Kib\'{e}di, G.~D. Dracoulis, A.~P. Byrne,
  \href{http://www.sciencedirect.com/science/article/pii/016890029090294G}{Lens-mode
  operation of a superconducting electron spectrometer in (hi, xn) reactions},
  Nuclear Instruments and Methods in Physics Research Section A: Accelerators,
  Spectrometers, Detectors and Associated Equipment 294~(3) (1990) 523 -- 533.
\newblock \href
  {http://dx.doi.org/http://dx.doi.org/10.1016/0168-9002(90)90294-G}
  {\path{doi:http://dx.doi.org/10.1016/0168-9002(90)90294-G}}.
\newline\urlprefix\url{http://www.sciencedirect.com/science/article/pii/016890029090294G}

\bibitem{butler_nima381}
P.~A. Butler, P.~M. Jones, K.~J. Cann, J.~F.~C. Cocks, G.~D. Jones, R.~Julin,
  W.~H. Trzaska, Electron spectroscopy using a multi-detector array, Nuclear
  Instruments and Methods in Physics Research Section A: Accelerators,
  Spectrometers, Detectors and Associated Equipment 381~(2–3) (1996) 433 --
  442.

\bibitem{Pakarinen2014}
J.~Pakarinen, P.~Papadakis, J.~Sorri, R.~D. Herzberg, P.~T. Greenlees, P.~A.
  Butler, P.~J. Coleman-Smith, D.~M. Cox, J.~R. Cresswell, P.~Jones, R.~Julin,
  J.~Konki, I.~H. Lazarus, S.~C. Letts, A.~Mistry, R.~D. Page, E.~Parr,
  V.~F.~E. Pucknell, P.~Rahkila, J.~Sampson, M.~Sandzelius, D.~A. Seddon,
  J.~Simpson, J.~Thornhill, D.~Wells,
  \href{https://doi.org/10.1140/epja/i2014-14053-6}{The {SAGE} spectrometer},
  The European Physical Journal A 50~(3) (2014) 53.
\newblock \href {http://dx.doi.org/10.1140/epja/i2014-14053-6}
  {\path{doi:10.1140/epja/i2014-14053-6}}.
\newline\urlprefix\url{https://doi.org/10.1140/epja/i2014-14053-6}

\bibitem{vanKlinken_nim98}
J.~V. Klinken, K.~Wisshak, Conversion electrons separated from high background,
  Nuclear Instruments and Methods 98~(1) (1972) 1 -- 8.

\bibitem{metlay_nima336}
M.~P. Metlay, J.~X. Saladin, I.~Y. Lee, O.~Dietzsch, The {ICEBall}: a multiple
  element array for in-beam internal conversion electron spectroscopy, Nuclear
  Instruments and Methods in Physics Research Section A: Accelerators,
  Spectrometers, Detectors and Associated Equipment 336~(1–2) (1993) 162 --
  170.

\bibitem{aengenvoort_epja1}
B.~Aengenvoort, W.~Korten, H.~H\"ubel, S.~Chmel, A.~G\"orgen, U.~J. van
  Severen, W.~Pohler, R.~Zinken, T.~H\"artlein, C.~Ender, F.~K\"ock, P.~Reiter,
  D.~Schwalm, F.~Schindler, J.~Gerl, R.~Schubart, F.~Azaiez, S.~Bouneau,
  J.~Duprat, I.~Deloncle, Conversion-electron $\gamma$-ray coincidence
  spectroscopy of superdeformed $^{135}\mathrm{Nd}$, The European Physical
  Journal A - Hadrons and Nuclei 1 (1998) 359--364.

\bibitem{Perkowski2014}
J.~Perkowski, J.~Andrzejewski, {\L}.~Janiak, J.~Samorajczyk, T.~Abraham,
  C.~Droste, E.~Grodner, K.~Hady\'nska-Kl\c{e}k, M.~Kisieli\'{n}ski,
  M.~Komorowska, M.~Kowalczyk, J.~Kownacki, J.~Mierzejewski, P.~Napiorkowski,
  A.~Korman, J.~Srebrny, A.~Stolarz, M.~Zieli\'{n}ska, {University of Lodz an
  electron spectrometer - A new conversion-electron spectrometer for
  ``in-beam'' measurements}, Review of Scientific Instruments 85~(043303).
\newblock \href {http://dx.doi.org/10.1063/1.4870899}
  {\path{doi:10.1063/1.4870899}}.

\bibitem{Battaglia2016}
A.~Battaglia, W.~Tan, R.~Avetisyan, C.~Casarella, A.~Gyurijinyan, K.~V.
  Manukyan, S.~T. Marley, A.~Nystrom, N.~Paul, K.~Siegl, K.~Smith, M.~K. Smith,
  S.~Y. Strauss, A.~Aprahamian,
  \href{https://doi.org/10.1140/epja/i2016-16126-x}{Measurements of conversion
  electrons in the s-process branching point nucleus {176Lu}}, The European
  Physical Journal A 52~(5) (2016) 126.
\newblock \href {http://dx.doi.org/10.1140/epja/i2016-16126-x}
  {\path{doi:10.1140/epja/i2016-16126-x}}.
\newline\urlprefix\url{https://doi.org/10.1140/epja/i2016-16126-x}

\bibitem{Papadakis2018}
P.~Papadakis, D.~M. Cox, G.~G. O'Neill, M.~J.~G. Borge, P.~A. Butler, L.~P.
  Gaffney, P.~T. Greenlees, R.~D. Herzberg, A.~Illana, D.~T. Joss, J.~Konki,
  T.~Kr{\"o}ll, J.~Ojala, R.~D. Page, P.~Rahkila, K.~Ranttila, J.~Thornhill,
  J.~Tuunanen, P.~Van~Duppen, N.~Warr, J.~Pakarinen,
  \href{https://doi.org/10.1140/epja/i2018-12474-9}{The {SPEDE} spectrometer},
  The European Physical Journal A 54~(3) (2018) 42.
\newblock \href {http://dx.doi.org/10.1140/epja/i2018-12474-9}
  {\path{doi:10.1140/epja/i2018-12474-9}}.
\newline\urlprefix\url{https://doi.org/10.1140/epja/i2018-12474-9}

\bibitem{ketelhut_nim214}
S.~Ketelhut, L.~J. Evitts, A.~B. Garnsworthy, C.~Bolton, G.~C. Ball,
  R.~Churchman, R.~Dunlop, G.~Hackman, R.~Henderson, M.~Moukaddam, E.~T. Rand,
  C.~E. Svensson, J.~Witmer,
  \href{http://www.sciencedirect.com/science/article/pii/S0168900214002708}{Simulated
  performance of the in-beam conversion-electron spectrometer, {SPICE}},
  Nuclear Instruments and Methods in Physics Research Section A: Accelerators,
  Spectrometers, Detectors and Associated Equipment 753 (2014) 154 -- 163.
\newblock \href {http://dx.doi.org/https://doi.org/10.1016/j.nima.2014.03.001}
  {\path{doi:https://doi.org/10.1016/j.nima.2014.03.001}}.
\newline\urlprefix\url{http://www.sciencedirect.com/science/article/pii/S0168900214002708}

\bibitem{martin_ieee55}
J.~P. Martin, C.~Mercier, N.~Starinski, C.~J. Pearson, P.~A. Amaudruz, The
  {TIGRESS} daq/trigger system, Nuclear Science, IEEE Transactions on 55~(1)
  (2008) 84--90.

\bibitem{Micron}
\href{http://www.micronsemiconductor.co.uk/multi-element-detectors-double-sided}{Micron
  semiconductor ltd.}
\newline\urlprefix\url{http://www.micronsemiconductor.co.uk/multi-element-detectors-double-sided}

\bibitem{SWAN}
\href{http://swanresearch.xorgate.com}{Swan research}.
\newline\urlprefix\url{http://swanresearch.xorgate.com}

\bibitem{Bildstein2015}
V.~Bildstein, P.~Garrett, S.~F. Ashley, G.~C. Ball, L.~Bianco,
  D.~Bandyopadhyay, J.~Bangay, B.~P. Crider, G.~Demand, G.~Deng, I.~Dillmann,
  A.~Finlay, A.~B. Garnsworthy, G.~Hackman, B.~Hadinia, R.~Kr{\" u}cken, K.~G.
  Leach, J.-P. Martin, M.~T. McEllistrem, C.~J. Pearson, E.~E. Peters, F.~M.
  Prados-Est{\' e}vez, A.~Radich, F.~Sarazin, C.~Sumithrarachchi, C.~E.
  Svensson, J.~R. Vanhoy, J.~Wong, S.~W. Yates, {DESCANT} and $\beta$-delayed
  neutron measurements at {TRIUMF}, EPJ Web of Conferences 93 (2015) 07005.

\bibitem{Laxdal2014}
R.~E. Laxdal, M.~Marchetto,
  \href{https://doi.org/10.1007/s10751-013-0884-8}{The {ISAC}
  post-accelerator}, Hyperfine Interactions 225~(1) (2014) 79--97.
\newblock \href {http://dx.doi.org/10.1007/s10751-013-0884-8}
  {\path{doi:10.1007/s10751-013-0884-8}}.
\newline\urlprefix\url{https://doi.org/10.1007/s10751-013-0884-8}

\bibitem{comsol_2012}
\href{http://www.comsol.com}{{COMSOL} multiphysics}.
\newline\urlprefix\url{http://www.comsol.com}

\bibitem{ALLISON2016186}
J.~Allison, K.~Amako, J.~Apostolakis, P.~Arce, M.~Asai, T.~Aso, E.~Bagli,
  A.~Bagulya, S.~Banerjee, G.~Barrand, B.~Beck, A.~Bogdanov, D.~Brandt,
  J.~Brown, H.~Burkhardt, P.~Canal, D.~Cano-Ott, S.~Chauvie, K.~Cho,
  G.~Cirrone, G.~Cooperman, M.~Cortés-Giraldo, G.~Cosmo, G.~Cuttone,
  G.~Depaola, L.~Desorgher, X.~Dong, A.~Dotti, V.~Elvira, G.~Folger,
  Z.~Francis, A.~Galoyan, L.~Garnier, M.~Gayer, K.~Genser, V.~Grichine,
  S.~Guatelli, P.~Guèye, P.~Gumplinger, A.~Howard, I.~Hřivnáčová,
  S.~Hwang, S.~Incerti, A.~Ivanchenko, V.~Ivanchenko, F.~Jones, S.~Jun,
  P.~Kaitaniemi, N.~Karakatsanis, M.~Karamitros, M.~Kelsey, A.~Kimura, T.~Koi,
  H.~Kurashige, A.~Lechner, S.~Lee, F.~Longo, M.~Maire, D.~Mancusi, A.~Mantero,
  E.~Mendoza, B.~Morgan, K.~Murakami, T.~Nikitina, L.~Pandola, P.~Paprocki,
  J.~Perl, I.~Petrović, M.~Pia, W.~Pokorski, J.~Quesada, M.~Raine, M.~Reis,
  A.~Ribon, A.~R. Fira, F.~Romano, G.~Russo, G.~Santin, T.~Sasaki, D.~Sawkey,
  J.~Shin, I.~Strakovsky, A.~Taborda, S.~Tanaka, B.~Tomé, T.~Toshito, H.~Tran,
  P.~Truscott, L.~Urban, V.~Uzhinsky, J.~Verbeke, M.~Verderi, B.~Wendt,
  H.~Wenzel, D.~Wright, D.~Wright, T.~Yamashita, J.~Yarba, H.~Yoshida,
  \href{http://www.sciencedirect.com/science/article/pii/S0168900216306957}{Recent
  developments in geant4}, Nuclear Instruments and Methods in Physics Research
  Section A: Accelerators, Spectrometers, Detectors and Associated Equipment
  835 (2016) 186 -- 225.
\newblock \href {http://dx.doi.org/https://doi.org/10.1016/j.nima.2016.06.125}
  {\path{doi:https://doi.org/10.1016/j.nima.2016.06.125}}.
\newline\urlprefix\url{http://www.sciencedirect.com/science/article/pii/S0168900216306957}

\bibitem{Kibedi2008}
T.~Kib{\' e}di, T.~W. Burrows, M.~B. Trzhaskovskaya, P.~M. Davidson, C.~W.
  Nestor,
  \href{http://www.sciencedirect.com/science/article/pii/S0168900208002520}{Evaluation
  of theoretical conversion coefficients using {BrIcc}}, Nuclear Instruments
  and Methods in Physics Research Section A: Accelerators, Spectrometers,
  Detectors and Associated Equipment 589~(2) (2008) 202 -- 229.
\newblock \href {http://dx.doi.org/https://doi.org/10.1016/j.nima.2008.02.051}
  {\path{doi:https://doi.org/10.1016/j.nima.2008.02.051}}.
\newline\urlprefix\url{http://www.sciencedirect.com/science/article/pii/S0168900208002520}

\end{thebibliography}

\end{document}